\documentclass[nofootinbib,twocolumn,showpacs,preprintnumbers,pre,aps]{revtex4-1}

\usepackage{graphicx}
\usepackage{bm}
\usepackage{amsmath}
\usepackage{amssymb}
\usepackage{color}

\begin{document}

\title{Pattern formation of skin cancers: Effects of cancer proliferation \\ 
and hydrodynamic interactions}%

\author{Takuma Hoshino}
\affiliation{
Department of Chemistry, Graduate School of Science, 
Tokyo Metropolitan University, Tokyo 192-0397, Japan}

\author{Ming-Wei Liu}
\affiliation{
Department of Physics, National Tsing Hua University, Hsinchu 30013, Taiwan}

\author{Kuo-An Wu}
\affiliation{
Department of Physics, National Tsing Hua University, Hsinchu 30013, Taiwan}

\author{Hsuan-Yi Chen}
\affiliation{
Department of Physics, National Central University, Jhongli 32001, Taiwan \\
Institute of Physics, Academia Sinica, Taipei 11529, Taiwan}

\author{Tatsuaki Tsuruyama}
\affiliation{Center for Anatomical Studies, Kyoto University Graduate School of Medicine, 
Kyoto 606-8501, Japan}

\author{Shigeyuki Komura}\email{komura@tmu.ac.jp}
\affiliation{
Department of Chemistry, Graduate School of Science, 
Tokyo Metropolitan University, Tokyo 192-0397, Japan}


\begin{abstract}
We study pattern formation of skin cancers 
by means of numerical simulation of a binary system consisting of cancer and healthy cells. 
We extend the conventional Model H for macrophase separations by considering a logistic growth 
of cancer cells and also a mechanical friction between dermis and epidermis. 
Importantly, our model exhibits a microphase separation due to the proliferation of cancer cells.
By numerically solving the time evolution equations of the cancer composition and its velocity, we 
show that the phase separation kinetics strongly depends on the cell proliferation rate as well as on 
the strength of hydrodynamic interactions.
A steady state diagram of cancer patterns is established in terms of these two dynamical parameters
and some of the patterns correspond to clinically observed cancer patterns.
Furthermore, we examine in detail the time evolution of the average composition of cancer cells and 
the characteristic length of the microstructures. 
Our results demonstrate that different sequence of cancer patterns can be obtained by changing 
the proliferation rate and/or hydrodynamic interactions.
\end{abstract}

\maketitle

\section{Introduction}
\label{introduction}

Tissue morphogenesis is a process in which multicellular organisms are dynamically formed 
in a coherent manner~\cite{MorphogenesisBook}.
Several deterministic and stochastic models to describe tissue regeneration using such as stem cells 
have been proposed from a theoretical point of view~\cite{Komarova2005,Pazdziorek2014}.
Recently, various analogies between viscoelastic fluids and biological tissues have 
been pointed out to investigate mechanical response of a biological tissue to an applied 
force~\cite{Basan2009,Ranft2010,Basan11,Brochard2012}.
Needless to say, studies on tumor dynamics are directly connected with medical diagnosis
and there have been many attempts to understand cancer behaviors across multiple biological 
scales~\cite{Kumar2009,Deisboeck2011,Rodriguez2013,Liedekerke2015}.
Although some correlations between cancer patterns and their malignancies are realized, it is 
not well-understood why and how such malignant patterns appear in tissues.
For example, a skin cancer called melanoma often exhibits characteristic surface patterns which 
are diagnosed by medical doctors~\cite{Chatelain2015}.
However, fundamental mechanisms that underlie such a pattern formation need to be further 
investigated.

Recently, some dynamical studies on skin lesions have been performed to discuss the morphological 
changes in early melanoma development by using a phase separation 
model~\cite{Chatelain2011NJP,Chatelain2011JTB,Balois2014,Balois2014JRSI}. 
Among these works, Chatelain \textit{et al.}\ investigated a binary system composed of cancer 
and healthy cells.
They demonstrated that not only the cell-cell adhesion but also the coupling to the diffusion of 
nutrients (oxygen) leads to the microstructure (e.g.\ ``dots" and ``nests") formation in the early 
stage melanoma~\cite{Chatelain2011NJP,Chatelain2011JTB,Balois2014}. 
These microstructures are analogous to those in block copolymer systems~\cite{HamleyBook}. 
In the model by Chatelain \textit{et al.}, the domain coarsening takes place due to diffusion 
process whereas hydrodynamic interactions are not considered.
Hence their model  can be regarded as an extension of  
``Model B"~\cite{Hohenberg77,LubenskyBook,OnukiBook} to take into account the formation 
of microstructures.
{
For bacterial colonies without hydrodynamic interactions, an arrested phase separation was 
explained only by considering a local density-dependent motility and the birth/death of 
bacteria~\cite{Cates2010}.}

In general, a biological tissue can be regarded as a viscoelastic material because it responds like a 
solid with finite elasticity at short time scales and behaves like a fluid with an effective viscosity 
at long time scales~\cite{Basan2009,Ranft2010,Basan11,Brochard2012}.
Since the ``differential adhesion hypothesis" was proposed by 
Steinberg~\cite{Steinberg63,Steinberg96}, the similarities between tissues and liquids have been 
recognized for a long time.
For example, by using particle tracking velocimetry in gastrulating \textit{Drosophila} embryos,
it was shown that cytoplasmic redistribution during ventral furrow formation is described
by the presence of hydrodynamic flows~\cite{He2014}.
In a recent study of tissue dynamics of a stratified epithelium, it was shown that a steady 
hydrodynamic flow of stratified epithelium is controlled by the cell proliferation rate~\cite{WeiTing2016,WeiTing2018}.
Although these works highlight the importance of liquid flows in the tissue dynamics, the effects of 
hydrodynamic interactions on the skin cancer dynamics have not been considered so far.

For ordinary fluid mixtures, on the other hand, it is well-known that hydrodynamic interactions play 
crucial roles in their phase separation dynamics.
This is because the convection of the composition field kinetically enhances the phase separation
in the presence of flows.
The standard model that takes into account the hydrodynamic effects is called ``Model H" that has 
been extensively studied in the literature~\cite{Hohenberg77,LubenskyBook,OnukiBook}.
For ordinary 3D fluid mixtures, Model H predicts that the domain size increases linearly with 
time~\cite{Siggia79,Kendon01}. 
This is much faster than the Brownian coagulation process~\cite{Binder74} or the Lifshitz-Slyozov 
evaporation-condensation process~\cite{LifshitzBook}.

In this paper, we study the pattern formation of skin cancers by means of numerical simulation 
of a binary system composed of cancer and healthy cells. 
Our main focus is to investigate the effects of cancer proliferation and hydrodynamic 
interactions on the phase separation kinetics. 
For this purpose, we shall extend the conventional Model H by incorporating a logistic growth
of cancer cells and a mechanical friction between dermis and epidermis. 
Similar to chemically reactive binary fluid mixtures~\cite{Glotzer94,Christensen96} or block 
copolymer melts~\cite{Oono88,Liu89,Bahiana90}, our model also exhibits a microphase separation 
due to the proliferation of cancer cells.

Performing numerical simulations of the time evolution of the cancer cell composition and the velocity field, 
we show that the phase separation dynamics is strongly affected by the cell proliferation rate as well 
as by the strength of hydrodynamic interactions.
We shall examine in detail how the average composition of cancer cells and the characteristic size 
of microstructures depend on these dynamical parameters. 
Our results also demonstrate that different sequence of cancer patterns can be obtained by changing 
the cancer proliferation rate and/or the hydrodynamic effects.  
Furthermore, our model can reproduce some of the clinically observed microstructures in melanoma.

\begin{figure}[tbh]
\centering
\includegraphics[scale=0.4]{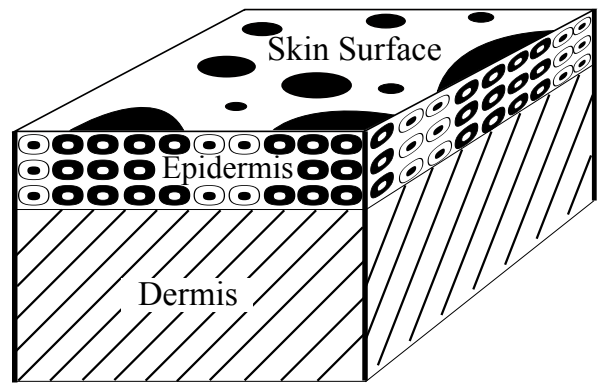}
\caption{
Schematic illustration of an epidermal tissue on dermis. 
The cell layer is assumed to be thin enough so that it can be regarded as a 2D fluid with 
hydrodynamic flows.
The fluid sheet is infinitely large  and we do not consider any out-of-plane deformation 
of the epidermal layer. 
The cell layer is composed of cancer cells (shown in black) and healthy cells (shown in white), 
and their areal compositions are defined by $\phi$ and $\psi$, respectively.
The two types of cell fill all the available space and satisfy the saturation constraint, i.e.,
$\phi+\psi=1$.
Further, the local velocities are denoted by $\mathbf{v}_{\phi}$ 
and $\mathbf{v}_{\psi}$ for cancer and healthy cells, respectively.
We also take into account a mechanical friction between dermis and epidermis that is characterized 
by the friction coefficient $\zeta$.
}
\label{fig1}
\end{figure}

In the next Section, we discuss the dynamical equations of a binary cell system in the 
presence of hydrodynamic interactions.
In Sec.~\ref{sec:result}, we present our simulation results for different proliferation rates and 
friction coefficients, and summarize them in terms of a steady state diagram as a function of 
these parameters. 
For qualitative arguments, we further perform structure analysis of the obtained patterns and 
give a scaling argument for the observed microphase separation.
In Sec.~\ref{sec:hydro}, we discuss the mechanisms for pattern formation in the early and 
late stages by using the amplitude equations method and the sharp interface model, respectively.
Finally, the summary of our work and some discussions are given in Sec.~\ref{sec:discussion}.

\section{Model}
\label{sec:model}

\subsection{Continuity equations}

Let us consider an epidermal cell layer on dermis as schematically depicted in Fig.~\ref{fig1}.
The cell layer is assumed to be thin enough such that it can be regarded as a two-dimensional
(2D) system characterized by a 2D vector $\mathbf{r}=(x,y)$. 
Here we do not consider any out-of-plane deformation of the epidermal layer. 
We assume that the cell layer is composed of cancer cells and healthy cells whose 
area fractions are denoted by $\phi(\mathbf{r},t)$ and $\psi(\mathbf{r},t)$
($0 \le \phi \le 1$ and $0 \le \psi \le 1$), respectively, which depend on time $t$.
For the hydrodynamic description, we define the corresponding local velocities by 
$\mathbf{v}_{\phi}(\mathbf{r},t)$ and $\mathbf{v}_{\psi}(\mathbf{r},t)$ for 
cancer cells and healthy cells, respectively.
We further assume that the two types of cell fill all the available space and always satisfy the 
saturation constraint $\phi+\psi=1$ at every point. 
This saturation constraint leads to the following incompressibility condition
\begin{align}
\nabla\cdot\mathbf{v}=0, 
\label{incompressibility}
\end{align}
where we have introduced the local average velocity 
\begin{align}
\mathbf{v}=\phi\mathbf{v}_{\phi}+\psi\mathbf{v}_{\psi},
\end{align}
which is weighted by the respective area fractions.

In order to take into account the proliferation of cancer cells and the death of healthy cells 
simultaneously, we consider the following continuity equations that are consistent with the above 
incompressibility condition: 
\begin{align}
& \frac{\partial \phi}{\partial t} 
+\nabla\cdot(\phi\mathbf{v}_{\phi}) =\Gamma(\phi),
\label{eq:conservation_c}
\\
& \frac{\partial \psi}{\partial t} 
+\nabla\cdot(\psi\mathbf{v}_{\psi}) =-\Gamma(\phi),
\label{eq:conservation_psi}
\end{align}
where the function $\Gamma(\phi)$ represents the composition-dependent cancer proliferation 
rate of epidermal cells. 
Among various possibilities, we choose here the following logistic growth function:
\begin{align}
\Gamma(\phi)=\gamma \phi \left(1-\frac{\phi}{\phi_\infty} \right),
\label{eq_cancerization2}
\end{align}
where the coefficient $\gamma>0$ is the cancer proliferation rate in the epidermal layer. 
{
Such a logistic growth was considered before to describe the effects of birth
and death in bacterial colonies~\cite{Cates2010}.}
Starting from an initial average composition, $\phi_0$, the cancer cell composition evolves 
toward a higher composition, $\phi_\infty$, whose value is roughly given by 
$\phi_\infty \approx 0.6$ -- $0.8$ depending on the cancer cell type~\cite{Jain}. 
Since the function $\Gamma(\phi)$ is positive, cancer cells proliferate during the phase separation 
while healthy cells die out due to the invasion of increased cancer cells, as described by 
Eq.~(\ref{eq:conservation_psi}).
Since the time-evolution of healthy cells is simply given by $\psi(\mathbf{r},t)=1-\phi(\mathbf{r},t)$ 
due to the saturation condition, we shall only consider Eq.~(\ref{eq:conservation_c}) in the following discussion.

We note here that the above introduced functional form of the proliferation rate, $\Gamma(\phi)$, is 
analogous to that considered in the previous model~\cite{Chatelain2011NJP,Chatelain2011JTB,Balois2014,Balois2014JRSI} 
in which they also included the diffusion of nutrient concentration.
One can easily show that the form of Eq.~(\ref{eq_cancerization2}) can be obtained by simply assuming 
that the nutrient concentration decreases linearly with the cancer composition $\phi$.
For the purpose of clarifying the effects of cancer proliferation and hydrodynamic interactions, 
it is sufficient to consider the above sigmoidal growth without introducing any additional field variable.

It should be mentioned that the above logistic growth of cancer cells can also originate 
from the mechanical coupling between the net cell division rates and pressure~\cite{Basan2009}.
In general, the cell division rates depend on mechanical pressure~\cite{Montel11,Montel12,Alessandri13,Delarue13}
and are characterized by the homeostatic pressure, i.e., the pressure for which cell division and apoptosis 
balance and no net growth occurs.
Near the homeostatic state, we are allowed to expand both the pressure and the net cell division rate to 
linear order in density difference around the homeostatic density~\cite{Basan2009}.
Such an effect also leads to the growth term in Eq.~(\ref{eq_cancerization2}).

\subsection{Dynamical equations}

{
Next we consider the time evolution equations for $\phi$ and $\mathbf{v}$.
By extending the standard Model H for phase separations with hydrodynamic 
interactions~\cite{Hohenberg77,LubenskyBook,OnukiBook}, the dynamical equations that 
are consistent with Eq.~(\ref{eq:conservation_c}) can be given by 
\begin{align}
\frac{\partial \phi}{\partial t}& =-\nabla\cdot (\phi\mathbf{v})+ L \nabla^2 \mu +\Gamma(\phi),
\label{eq_phic}
\\
\rho \frac{\partial \mathbf{v}}{\partial t}& =
\eta\nabla^2\mathbf{v} -\nabla p +\nabla\cdot\boldsymbol{\Sigma}-\zeta\mathbf{v}, 
\label{eq_velocity}
\end{align}
together with the incompressibility condition in Eq.~(\ref{incompressibility}).
In the above equations, $L$ is the transport coefficient, $\mu$ is the chemical potential, 
$\rho$ is the mass density, $\eta$ is the viscosity, $p$ is the 2D pressure, 
$\boldsymbol{\Sigma}$ is the stress tensor due to the composition 
gradient, and $\zeta$ is the friction coefficient. 
For simplicity, we assume that both $\rho$ and $\eta$ are constants and do not depend on $\phi$.
Moreover, we consider the case when the transport coefficient $L$ is independent of 
$\phi$~\cite{Tiribocchi15}, because a composition dependent transport coefficient would not 
alter the asymptotic dynamics~\cite{Puri97,Ahluwalia99}. 
In the present work, we do not include any stochastic noise.
}

{
The last term $-\zeta\mathbf{v}$ in Eq.~(\ref{eq_velocity}) represents the frictional 
dissipation between the epidermal layer and dermis. 
In human tissues, such a friction arises from the adhesion of integrins that connect a keratin intracellular 
network to collagen fibers of basement membranes.
With this frictional term, the total momentum is no longer conserved within the 2D fluid sheet.  
Furthermore, the friction coefficient $\zeta$ controls the strength of hydrodynamic interactions.
Namely, hydrodynamics does not play any role when $\zeta \rightarrow \infty$, whereas 
hydrodynamic interactions are fully present when $\zeta \rightarrow 0$.
Later we shall systematically change the value of $\zeta$ to investigate the effects of hydrodynamic 
interactions on the phase separation kinetics.
}

{
To further obtain the chemical potential $\mu$ and the stress tensor $\boldsymbol{\Sigma}$,
we introduce the total free energy describing the phase separation of a cell mixture. 
Following Wise \textit{et al.}\ who discussed a continuum model of multi-species tumor 
growth~\cite{Wise10}, we use the following form for a binary cellular system:
\begin{align}
F = & \int d\mathbf{r} \, \biggl[ 
\frac{1}{a^2 \beta} \bigl[\phi\ln\phi+(1-\phi)\ln(1-\phi)
\nonumber\\
&+\chi\phi(1-\phi)\bigr]+\frac{\kappa}{2}(\nabla \phi)^{2} \biggr].
\label{freeenergy}
\end{align}
Here, $a$ has the dimension of length, $\beta^{-1}$ has the dimension of energy,
$\chi$ is a dimensionless interaction parameter between cancer and healthy cells, 
and $\kappa>0$ is a quantity related to the line tension in the 2D cellular sheet.
}

{
Since the above equation has the same form as the Flory-Huggins free energy,
a phase separation occurs for the condition $\chi > 2$~\cite{DoiBook}.
Notice that the local terms can be replaced by any other phenomenological description which 
exhibits a phase separation at sufficiently strong repulsion between the different cell types.
Hence the exact functional form is not important and different forms of free energy were
proposed in Refs.~\cite{Chatelain2011NJP,Chatelain2011JTB,Balois2014,Balois2014JRSI}.
}

{
The chemical potential $\mu$ is obtained from the functional derivative of the total free energy 
$F$ with respect to $\phi$~\cite{DoiBook}
\begin{align}
\mu=\frac{\delta F}{\delta\phi}=
\frac{1}{a^2 \beta} \left[ \ln\frac{\phi}{1-\phi}+\chi(1-2\phi) \right] -\kappa\nabla^{2}\phi.
\label{eq_chem}
\end{align}
On the other hand, the stress tensor due to the composition gradient $\boldsymbol{\Sigma}$ 
is given by~\cite{DoiBook} 
\begin{align}
\Sigma_{ij}=
-\kappa\frac{\partial \phi}{\partial r_i}\frac{\partial \phi}{\partial r_j}, 
\end{align}
with $i, j=x, y$.
}

{
The coupled Eqs.~(\ref{eq_phic}) and (\ref{eq_velocity}) together with the incompressibility condition
in Eq.~(\ref{incompressibility})  constitute our model for skin cancers and provide us with a new type 
of phase separation dynamics.
In the absence of the cancer proliferation effect, i.e., $\gamma=0$, the above model reduces
to conventional models for macrophase separations~\cite{Hohenberg77,LubenskyBook,OnukiBook}. 
When $\gamma=0$, our model reduces to Model H in the limit of $\zeta \rightarrow 0$ with full hydrodynamic 
interactions, while it corresponds to Model B in the limit of $\zeta \rightarrow \infty$ for which 
hydrodynamic interactions are completely suppressed.
The case of $\gamma \neq 0$ showing an arrested phase separation was studied for the pattern 
formation of bacterial colonies in the absence of hydrodynamic interactions~\cite{Cates2010}.
}

\subsection{Simulation method}

{
We numerically solve Eqs.~(\ref{incompressibility}), (\ref{eq_phic}) and (\ref{eq_velocity}) by using a standard 
Euler's method on a 2D square lattice of size $512\times 512$
with periodic boundary conditions.
The pressure field $p$ is calculated  with the marker-and-cell method in each time 
step~\cite{Harlow65}.
It is convenient to use the quantities $a$, $\beta^{-1}$,  and $a^4\beta/L$ 
to scale length, energy, and time, respectively.
The numerical estimations for these quantities will be discussed in Sec.~\ref{sec:discussion}.
Then the dimensionless velocity becomes 
$\widetilde{\mathbf{v}}\equiv (a^3\beta/L)\mathbf{v}$ and the dimensionless model parameters 
are defined by 
\begin{align}
\widetilde{\rho}\equiv\frac{L^2}{a^4\beta}\rho,~~~
\widetilde{\eta}\equiv\frac{ L}{a^2}\eta,~~~
\widetilde{\gamma}\equiv\frac{a^4\beta}{L}\gamma,~~~
\widetilde{\zeta}\equiv  L \zeta, ~~~
\widetilde{\kappa}\equiv \beta \kappa.
\label{dimensionless}
\end{align}
}

{
With the above rescaling, we end up with the following six dimensionless parameters: $\chi$, 
$\widetilde{\kappa}$, $\widetilde{\rho}$, $\widetilde{\eta}$, $\widetilde{\gamma}$, and $\widetilde{\zeta}$.
Among these parameters, we have fixed four of them as $\chi=2.5$, 
$\widetilde{\kappa}=1.0$, $\widetilde{\rho}=0.3$, and $\widetilde{\eta}=1.0$ in our simulations.
Moreover, the initial and the final values of the cancer area fractions are chosen as $\phi_0=0.3$ 
and $\phi_\infty=0.8$~\cite{Jain}, respectively.
In the following, we shall mainly vary the two dynamical parameters, $\widetilde{\gamma}$ and 
$\widetilde{\zeta}$, to see the effects of cancer proliferation and hydrodynamic interactions on the 
pattern formation of skin cancers.
Physically speaking, the strength of the hydrodynamic interaction should be characterized 
by a dimensionless number $\zeta a^2/\eta = \widetilde{\zeta}/\widetilde{\eta}$ that involves 
both the viscosity and the friction coefficient.
Since we set $\widetilde{\eta}=1.0$ in our simulations, the parameter $\widetilde{\zeta}$ controls 
the strength of the hydrodynamic interaction.
When we present the simulation results in Sec.~\ref{sec:result}, the above tilde notation is omitted 
and all the quantities are treated as dimensionless numbers.
}

\begin{figure*}[tb]
\centering
\includegraphics[width=0.95 \linewidth]{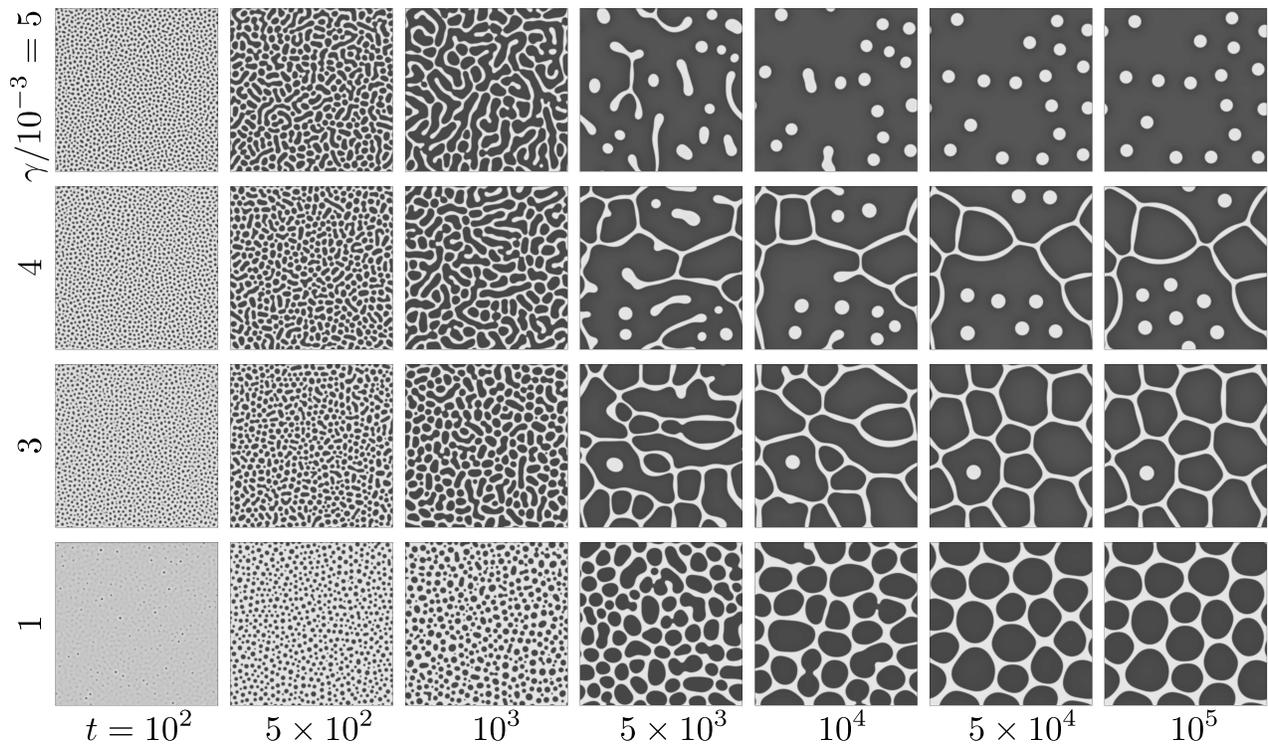}
\caption{
Time evolutions of cancer area fraction $\phi(\mathbf{r}, t)$ for four different values of the 
cancer proliferation rate $\gamma =1, 3, 4$ and $5 \times 10^{-3}$ (bottom to top) in the presence 
of full hydrodynamic interactions ($\zeta=0$). 
The other dimensionless parameters are $\phi_0=0.3$, $\phi_\infty=0.8$, $\chi=2.5$, $\kappa=1$, 
$\rho=0.3$
and $\eta=1.0$.
The system size is $512 \times 512$ and the velocity filed is not shown.
In the present greyscale representation, the values $0$ and $1$ correspond to white and black, respectively.
For $\gamma =1$ and $5 \times 10^{-3}$, see also \texttt{SM1.mp4} and \texttt{SM2.mp4}, respectively,
in the SM.
}
\label{patterns1}
\end{figure*}

\begin{figure*}[tb]
\centering
\includegraphics[width=0.95 \linewidth]{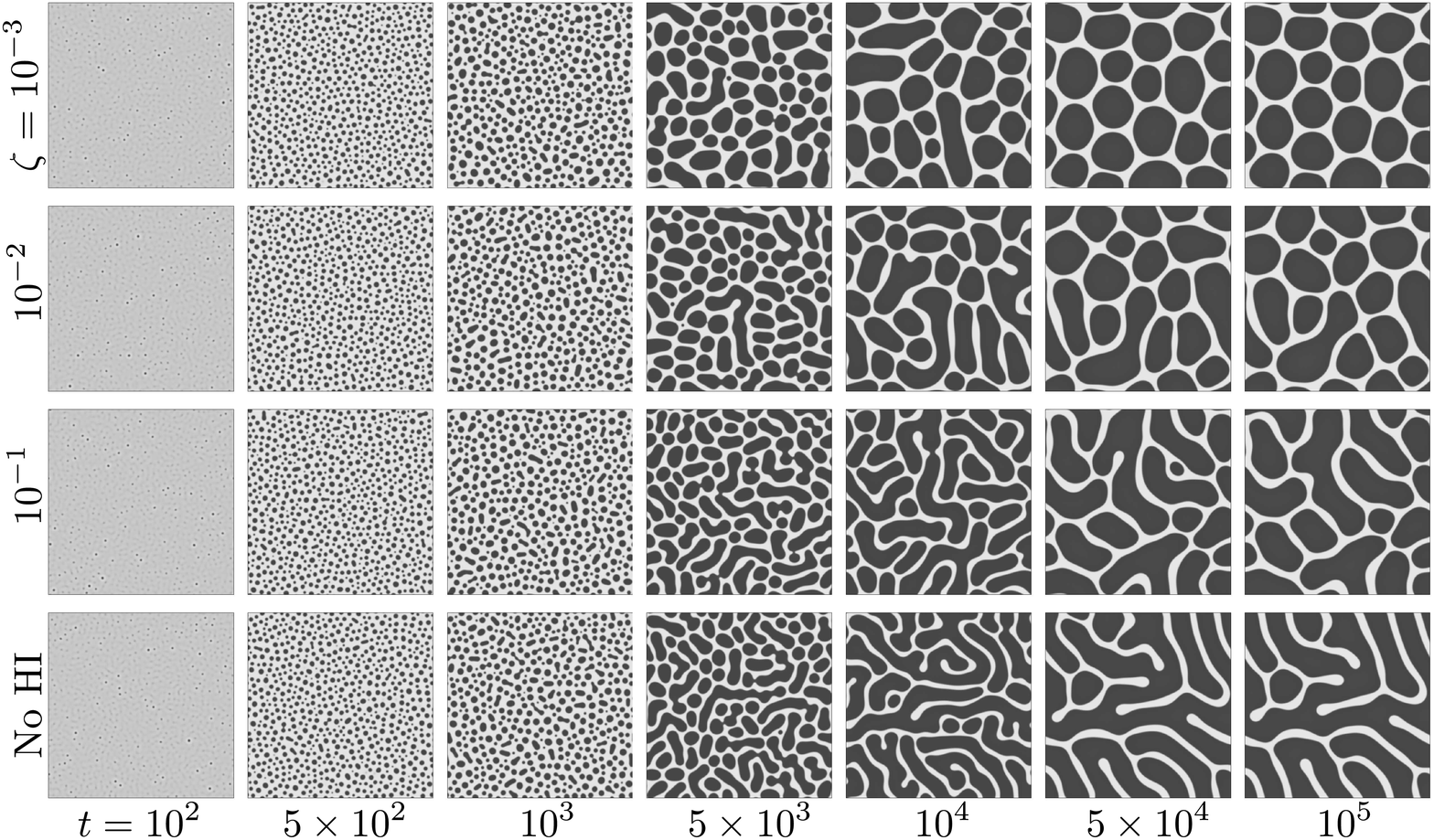}
\caption{
Time evolutions of cancer area fraction $\phi(\mathbf{r}, t)$ for four different values of the 
friction coefficient $\zeta=10^{-3}, 10^{-2}, 10^{-1}$ and $\infty$ (top to bottom) while 
the cancer proliferation rate is fixed to $\gamma=1\times 10^{-3}$. 
Notice that the limit $\zeta \rightarrow \infty$ is equivalent to the complete absence of hydrodynamic interactions
(No HI).
In practice, such a situation was simulated by omitting the advection term in Eq.~(\ref{eq_phic}).
The other parameters are the same as those in Fig.~\ref{patterns1}.
The values $0$ and $1$ correspond to white and black, respectively.
For No HI, see also \texttt{SM3.mp4} in the SM.} 
\label{patterns2}
\end{figure*}

\section{Simulation results}
\label{sec:result}

\subsection{Pattern formation dynamics}

In this Section, we present the results of the numerical simulations of the proposed model.
We first define the spatially averaged composition of cancer cells as 
\begin{align}
\langle \phi (t) \rangle = \frac{1}{A} \int d\mathbf{r} \, \phi(\mathbf{r},t),
\label{averagedensity}
\end{align}
where $A$ is the total area of the system. 
Because of the cancer proliferation,  $\langle \phi (t) \rangle$ varies from the initial value 
$\phi_0=0.3$ towards the stationary value $\phi_\infty=0.8$. 
Typical time evolutions of cancer patterns are shown in Fig.~\ref{patterns1} when 
$\zeta=0$ for four different values of the cancer proliferation rate $\gamma =1, 3, 4$ and 
$5 \times 10^{-3}$
(bottom to top).
Notice that $\zeta=0$ corresponds to the case with full hydrodynamic interactions.

\begin{figure}[tb]
\centering
\includegraphics[scale=0.35]{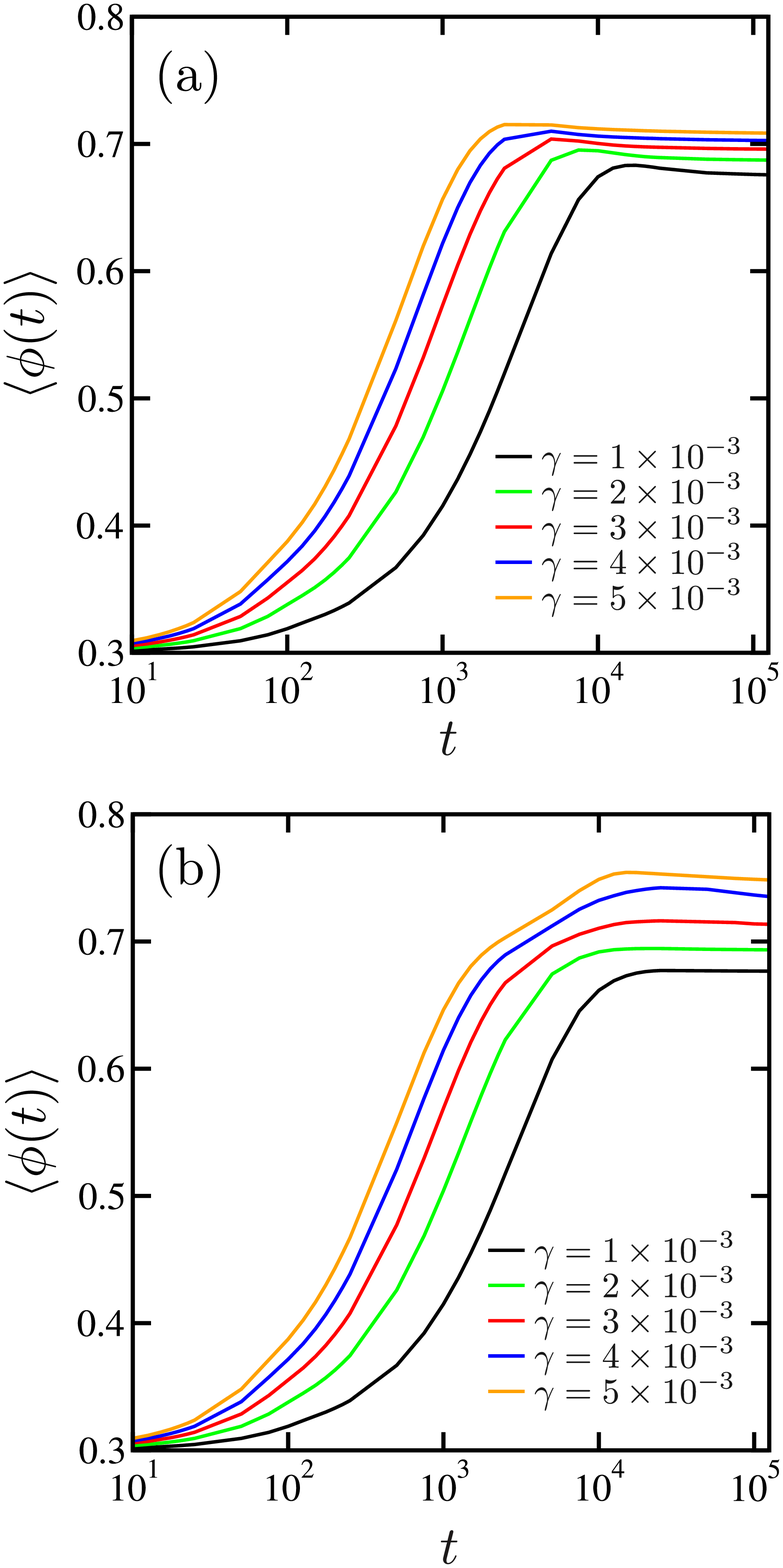}
\caption{
(Color Online) 
Plots of the average area fraction $\langle \phi (t) \rangle$, defined by 
Eq.~(\ref{averagedensity}), as a function of time $t$  
(a) in the absence of hydrodynamic interactions and 
(b) in the presence of full hydrodynamic interactions ($\zeta=0$).
In the former case, simulations were performed by omitting the advection term in Eq.~(\ref{eq_phic}).
The cancer proliferation rate is changed as $\gamma = 1, 2, 3, 4$ and $5 \times 10^{-3}$ (from bottom to top).
The other parameters are the same as those in Fig.~\ref{patterns1}.}
\label{phi-t}
\end{figure}

Let us first discuss the case of small proliferation rate $\gamma =1\times 10^{-3}$ 
(bottom panels in Fig.~\ref{patterns1} and the movie \texttt{SM1.mp4} in the SM).
In the initial stage at around $t=5\times 10^2$, dots of cancer cells (shown in black) are formed 
within a continuous healthy region (shown in white). 
We shall call such a structure as a ``cancer-in-healthy" (C/H) pattern.
As time evolves, smaller cancer domains collide and merge to form larger domains at around 
$t=10^4$.
However, not all the cancer domains are connected to each other even though $\langle \phi (t) \rangle$ 
already exceeds the critical composition $\phi_{\rm c}=0.5$.
The C/H pattern
in the late stage no longer evolves in time and the system attains a steady state without 
undergoing a macroscopic phase separation.
This result shows that our model exhibits a microphase separation.

When the cancer proliferation rate is larger such as when $\gamma =5 \times 10^{-3}$ (top panels 
in Fig.~\ref{patterns1} and the movie  \texttt{SM2.mp4}), healthy regions transform to cancer domains 
even in the early stage, and the C/H pattern is already formed at around $t=10^2$.  
As the average composition $\langle \phi (t) \rangle$ increases, a {locally} bicontinuous 
cancer structure is formed at around $t=10^3$. 
However, such a {locally} bicontinuous structure is destroyed later and smaller healthy domains emerge. 
At this stage, black cancer domains are almost fully connected to form a large continuous domain at 
around $t=5\times 10^3$.
In the late stage, circular domains of healthy cells appear in the network of cancer cells.
Such a structure will be called as a ``healthy-in-cancer" (H/C) pattern.
These circular healthy domains do not coarsen any more in the long time and result in a microphase 
separation.

When the proliferation rate is intermediate such as when $\gamma =3 \times 10^{-3}$, healthy domains 
are elongated and form a narrow continuous network.
Moreover, cancer domains in the late stage at around $t=10^5$ take polygonal shapes rather than 
circular shapes. 
For $\gamma =4 \times 10^{-3}$, a coexistence between the C/H and H/C patterns is observed 
as a steady state structure.

So far we have explained the effects of cancer proliferation rate $\gamma$ in the presence of full 
hydrodynamic interactions, i.e., $\zeta=0$.   
Next we investigate the hydrodynamic effects by changing the friction coefficient $\zeta$. 
In Fig.~\ref{patterns2}, we present the time evolutions of cancer patterns when the  
proliferation rate is fixed to $\gamma=1\times 10^{-3}$ while the friction coefficient is varied as 
$\zeta=10^{-3}, 10^{-2}, 10^{-1}$ and $\infty$ (top to bottom).
Notice that hydrodynamic interactions are completely absent when $\zeta \rightarrow \infty$.
In practice, this situation is simulated by omitting the advection term in Eq.~(\ref{eq_phic}) 
which is then decoupled from the Stokes equation.
When the friction coefficient is small such as  when $\zeta = 10^{-3}$ (top panels in Fig.~\ref{patterns2}), 
the time evolution of cancer pattern is similar to that obtained with full hydrodynamic interactions (bottom 
panels in Fig.~\ref{patterns1}).
However, the steady state cancer domains at around $t=10^5$ are more elongated.
The appearance of elongated domains in the steady state is more remarkable for $\zeta = 10^{-2}$.

As the hydrodynamic interactions are further weakened such as when $\zeta=10^{-1}$, cancer domains 
are more elongated especially in the late stage patterns.
Here we emphasize again that the major cancer domains are disconnected while the minor healthy 
domains form a continuous network structure. 
When hydrodynamic interactions are completely absent (bottom panels in Fig.~\ref{patterns2} and 
the movie \texttt{SM3.mp4}), we eventually obtain an asymmetric bicontinuous (AB) structure 
{at least locally}. 
In this structure, both the wider interconnected cancer domain and the narrower interconnected 
healthy domain are convoluted to each other for $t \ge 10^4$.

\subsection{Average cancer composition}

In Fig.~\ref{phi-t}, we have plotted the average cancer composition $\langle \phi (t)\rangle$, 
defined by Eq.~(\ref{averagedensity}), as a function of time $t$ by changing the cancer proliferation 
rate $\gamma$.
To calculate this quantity, average over five independent runs (starting from different initial 
configurations) has been taken.
Figure~\ref{phi-t}(a) is the case when hydrodynamic interactions are completely absent.
As $\gamma$ is increased, the saturation time becomes smaller and the saturated value of 
$\langle \phi (t) \rangle$ becomes larger. 
It is interesting to note that $\langle \phi (t) \rangle$ overshoots before it reaches the stationary value.

When hydrodynamic interactions are fully present ($\zeta=0$), on the other hand, the time evolutions 
of $\langle \phi (t) \rangle$ are different as presented in Fig.~\ref{phi-t}(b).
Here we notice that the value of $\langle \phi (t) \rangle$ becomes slightly larger when the hydrodynamic 
interactions are present especially for larger $\gamma$ values.
However, the overshooting behavior of $\langle \phi (t) \rangle$ is suppressed in Fig.~\ref{phi-t}(b).
These results indicate that hydrodynamic interactions affect not only the steady state behavior 
but also the transient dynamics of pattern formation.

\subsection{Steady state diagram}

\begin{figure}[tb]
\centering
\includegraphics[scale=0.4]{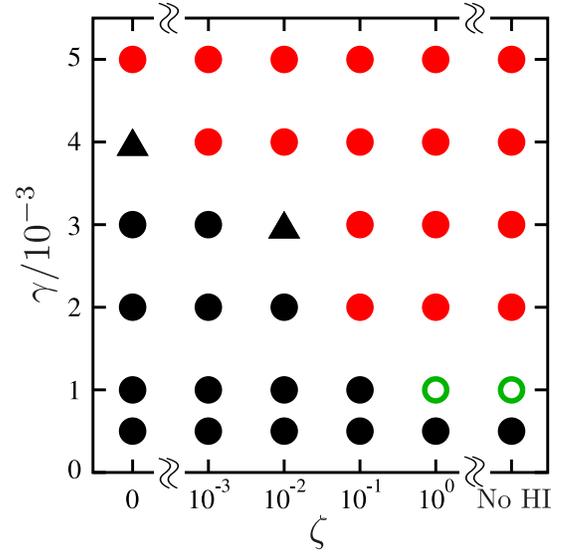}
\caption{
(Color Online) Steady state diagram of cancer patterns obtained for different cancer proliferation rate 
$\gamma$ and friction coefficient $\zeta$ (controlling the strength of hydrodynamic interactions). 
The other parameters are the same as those in Fig.~\ref{patterns1}.
Hydrodynamic interactions are fully present when $\zeta=0$, whereas they are completely 
absent in the limit of $\zeta \rightarrow \infty$ (No HI). 
The latter situation was simulated by omitting the advection term in Eq.~(\ref{eq_phic}).
Black circles correspond to cancer-in-healthy (C/H) patterns 
(such as the bottom right pattern in Fig.~\ref{patterns1}),  
red (light gray) circles correspond to healthy-in-cancer (H/C) patterns 
(such as the top right pattern in Fig.~\ref{patterns1}), and 
green (open) circles correspond to {(locally)} asymmetric bicontinuous (AB) patterns 
(such as the bottom right pattern in Fig.~\ref{patterns2}) in the respective steady states.
Black triangles indicate the coexistence between C/H and H/C patterns 
(such as $\gamma =4 \times 10^{-3}$ and $t=10^5$ in Fig.~\ref{patterns1}).}
\label{diagram}
\end{figure}

Next we have systematically varied the proliferation rate $\gamma$ and the friction coefficient $\zeta$  
to see how the steady state structures depend on these dynamic parameters.
We have mentioned before that there are at least three different steady state patterns:
cancer-in-healthy (C/H), healthy-in-cancer (H/C) and asymmetric bicontinuous (AB) patterns. 
The obtained steady state patterns are classified into these three cases for different combinations
of $\gamma$ and $\zeta$.
In Fig.~\ref{diagram}, we summarize the results in terms of a steady state diagram in which  
the three different cases, C/H, H/C, and AB are distinguished.
The triangle indicates the coexistence between C/H and H/C patterns.

The C/H pattern clinically corresponds to the globule pattern of melanoma, and is typically observed 
when the proliferation rate $\gamma$ is small and hydrodynamic interactions are strong (small $\zeta$).
The AB pattern appears when hydrodynamic interactions are weak or fully suppressed (large $\zeta$) 
while the proliferation rate $\gamma$ is relatively small.
The AB pattern may correspond to the stripe pattern of melanoma mainly found in human 
palms or soles.
Finally, the H/C pattern typically appears when both $\gamma$ and $\zeta$ are large. 
When the proliferation rate is as large as $\gamma=5 \times 10^{-3}$, only the H/C pattern is obtained 
irrespective of the strength of hydrodynamic interactions.
In contrast to the other two cases, however, the H/C pattern is usually not diagnosed in typical 
skin cancers because domains of healthy cells are completely destroyed by invasive cancer cells.

In the case of an ordinary microphase separation, the late stage structure should be the H/C pattern
when $\langle \phi (t) \rangle > 0.5$.
As shown in Fig.~\ref{diagram}, however, we obtain either the C/H pattern or the AB pattern for 
different combinations of $\zeta$ and $\gamma$, especially when $\gamma$ is small. 
This is one of the unique features of the proposed model for cancer cells with hydrodynamic 
interactions. 
Since these steady state patterns are typically obtained in the presence of hydrodynamic 
interactions, we consider that they  appear kinetically and do not correspond to equilibrium 
microstructures.
The different mechanisms for the pattern formation will be discussed in 
Sec.~\ref{sec:hydro}.

\subsection{Structure analysis}

\begin{figure}[tb]
\centering
\includegraphics[scale=0.35]{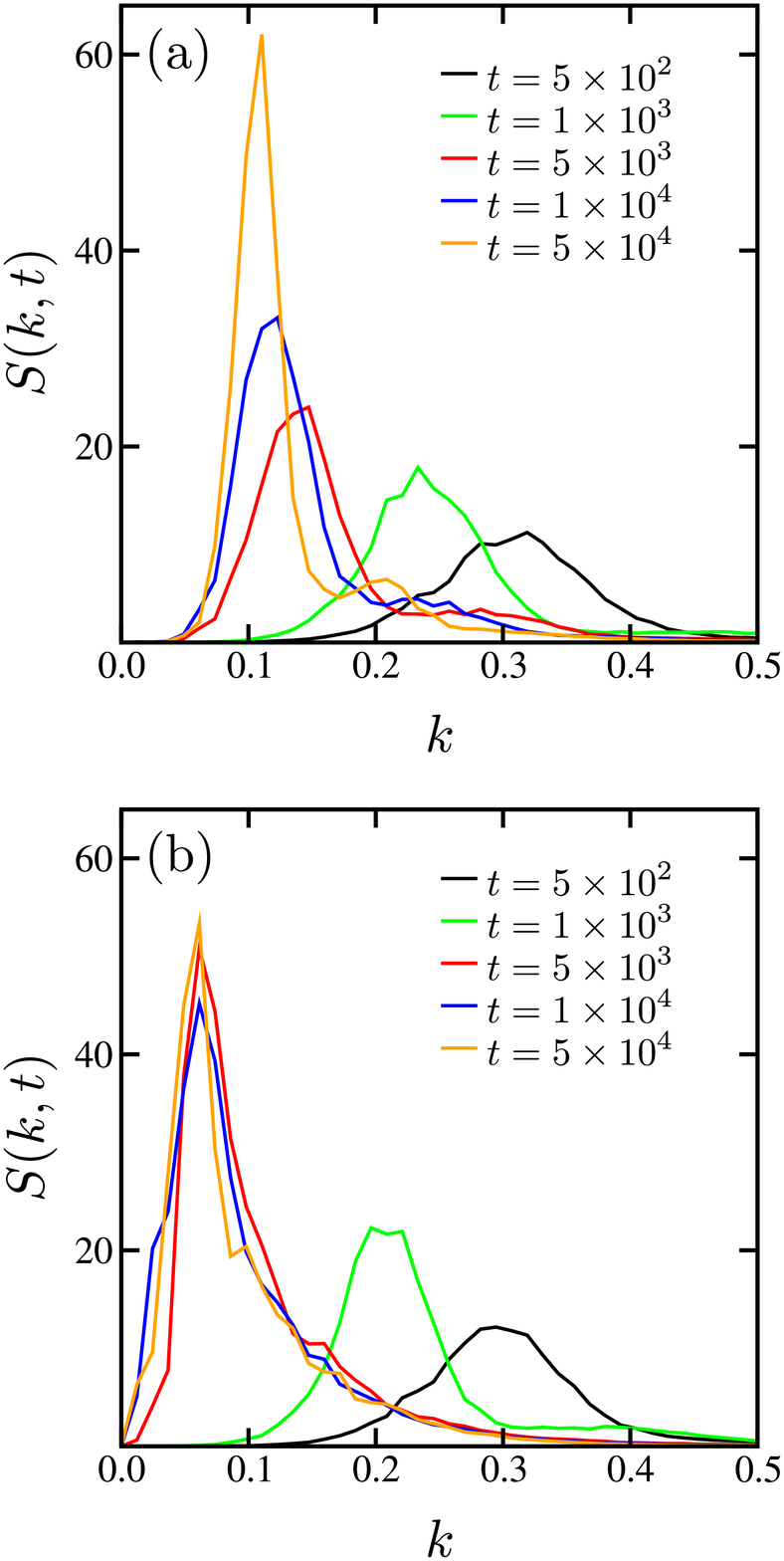}
\caption{
(Color Online) Plots of the circularly averaged structure factor $S(k,t)$ as a function of 
the wave number $k$ for different time steps $t$ (from right to left)
(a) in the absence of hydrodynamic interactions and 
(b) in the presence of full hydrodynamic interactions ($\zeta=0$).
The cancer proliferation rate is fixed to $\gamma = 3\times 10^{-3}$, while 
the other parameters are the same as those in Fig.~\ref{patterns1}.
Notice that the real space pattern evolution that corresponds to (b) is presented in Fig.~\ref{patterns1}.}
\label{skt}
\end{figure}

\begin{figure}[tb]
\centering
\includegraphics[scale=0.35]{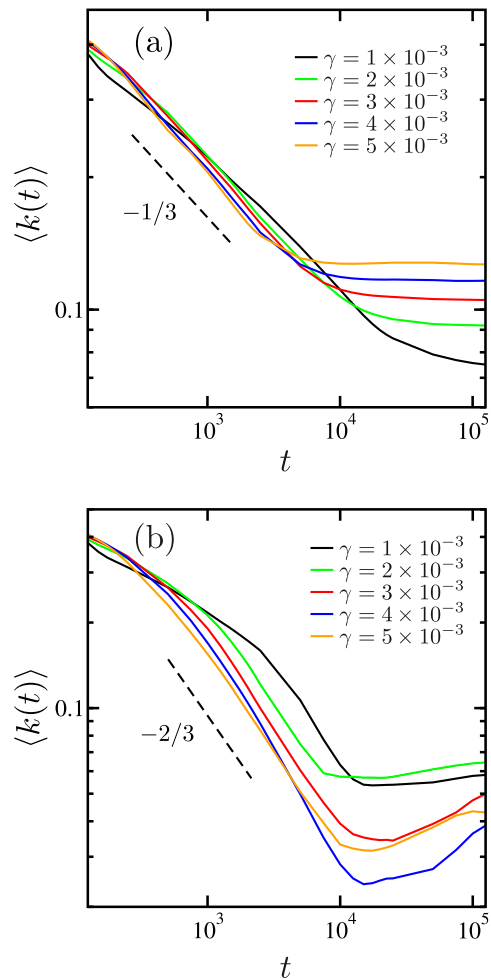}
\caption{
(Color Online) Log-log plots of the characteristic wave number $\langle k (t) \rangle$, defined by 
Eq.~(\ref{averagek}), as a function of time $t$ 
(a) in the absence of hydrodynamic interactions and 
(b) in the presence of full hydrodynamic interactions ($\zeta=0$).
The cancer proliferation rate is changed as $\gamma = 1, 2, 3, 4$ and $5 \times 10^{-3}$
(from bottom to top at $t=10^5$ in (a) and from right to left for the intermediate time region in (b)).
The other parameters are the same as those in Fig.~\ref{patterns1}.
The dashed lines indicate the power-law behaviors with the respective slopes $-1/3$
in (a) and $-2/3$ in (b).}
\label{k-t}
\end{figure}

To analyze the time evolutions of the patterns quantitatively, we have calculated  
their structure factors.
Let $\delta \phi(\mathbf{r},t)$ be the deviation of $\phi(\mathbf{r},t)$ from its average value,
$\delta \phi(\mathbf{r},t)=\phi(\mathbf{r},t)-\langle \phi (t) \rangle$, 
where $\langle \phi (t) \rangle$ defined in Eq.~(\ref{averagedensity}) depends on time.
First we introduce the spatial Fourier transform of $\delta \phi(\mathbf{r},t)$ by 
\begin{align}
\delta \phi_{\mathbf{k}}(t) = \int d\mathbf{r} \, \delta \phi(\mathbf{r},t) e^{-i \mathbf{k} \cdot \mathbf{r}},
\label{Fourier}
\end{align}
where $\mathbf{k}=(k_x,k_y)$ is a 2D wave vector. 
Then the structure factor is defined as 
\begin{align}
S(\mathbf{k},t)=\langle \delta\phi_{\mathbf{k}}(t) 
\delta \phi_{-\mathbf{k}}(t) \rangle,
\label{structure}
\end{align}
where the average is over the ensemble of systems.
Using the circularly averaged structure factor $S(k,t)$ with $k=\vert \mathbf{k} \vert$, 
we calculate the following (inverse) characteristic length scale of patterns~\cite{Shinozaki93} 
\begin{align}
\langle k (t) \rangle= \frac{\int dk \, k^{-1} S(k,t)}{\int dk \, k^{-2} S(k,t)},
\label{averagek}
\end{align}
where we omit $k=0$ in the integrals.

In Fig.~\ref{skt}, we plot the time evolutions of the circularly averaged structure factor $S(k,t)$
as a function of the wave number $k$ when $\gamma = 3\times 10^{-3}$.
Figure~\ref{skt}(a) corresponds to the case when hydrodynamic interactions are absent, while 
Fig.~\ref{skt}(b) presents the case with full hydrodynamics.  
By comparing these two cases, we see that the early stage structures are similar as long as the 
proliferation rate $\gamma$ is the same. 
In the intermediate stage, however, the microstructure formation  is faster in the presence of hydrodynamic 
interactions, and the peak position is shifted to a smaller $k$-value in Fig.~\ref{skt}(b).
We also find that the peak height in the late stage is slightly smaller in Fig.~\ref{skt}(b) than 
that in Fig.~\ref{skt}(a).

In Fig.~\ref{k-t}, we have plotted the characteristic wave number $\langle k (t) \rangle$, 
defined by Eq.~(\ref{averagek}), as a function of time.
The proliferation rate $\gamma$ is similarly changed as in Fig.~\ref{phi-t} and the 
average over five independent runs has been taken as before.
As shown in Fig.~\ref{k-t}(a) when hydrodynamic interactions are absent, the average wave 
number $\langle k (t) \rangle$ saturates at larger values (smaller structures) when $\gamma$ 
is increased.
This means that $\gamma$ is an important parameter that controls the characteristic length 
scale of the steady state microstructures. 
Comparing Figs.~\ref{phi-t}(a) and \ref{k-t}(a), we notice that the saturation times for 
$\langle \phi (t) \rangle$ roughly correspond to those for $\langle k (t) \rangle$.

The effects of hydrodynamic interactions on $\langle k (t) \rangle$ can be seen in Fig.~\ref{k-t}(b)
for which we have set $\zeta=0$. 
Here $\langle k (t) \rangle$ shows a large decrease up to the intermediate stage. 
This result indicates that hydrodynamic interactions tend to form larger domains even though they are
only transient structures. 
Interestingly, a minimum of $\langle k (t) \rangle$ appears at around $t=10^4$ and $\langle k (t) \rangle$
exhibits an undershooting behavior. 
Hence the transient domain size depends not only on the proliferation rate $\gamma$ 
but also on the friction coefficient $\zeta$.
According to Fig.~\ref{k-t}(b), however, the late stage dynamics of $\langle k (t) \rangle$ has not 
yet reached the steady state completely. 
Such a long-lived dynamics is also different from the case without hydrodynamic interactions.

\begin{figure}[tb]
\centering
\includegraphics[scale=0.35]{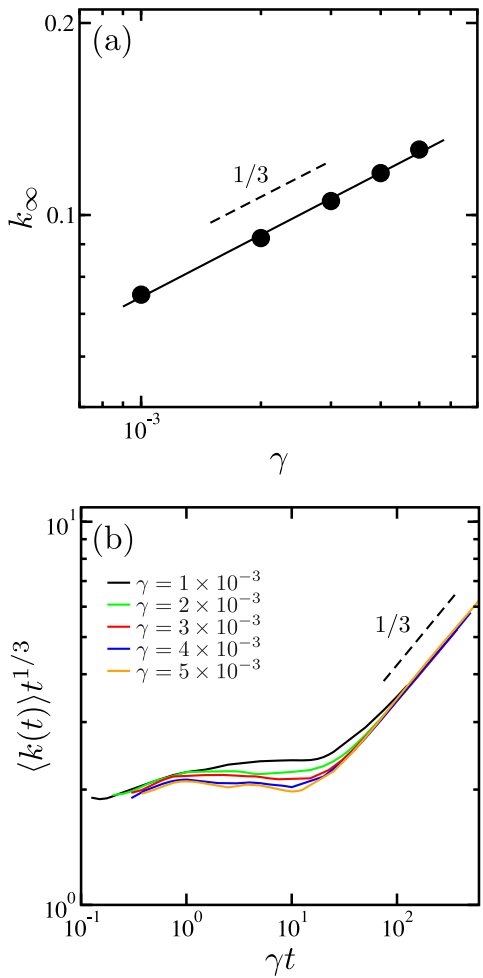}
\caption{
(Color Online) (a) Log-log plot of the steady state value of the characteristic wave number 
$k_\infty$ in  Fig.~\ref{k-t}(a) as a function of the proliferation rate $\gamma$ in the absence of 
hydrodynamic interactions.
From the slope of the fitted straight line, we find a power-law relation 
$k_\infty \sim \gamma^{0.32}$.
(b) Log-log plot of $\langle k (t) \rangle t^{1/3}$ as a function of the dimensionless variable $\gamma t$
using all the data in Fig.~\ref{k-t}(a).
The collapse of all the data confirms the validity of the scaling assumption in Eq.~(\ref{scaling}).
The dashed lines indicate the power-law behaviors with the respective slopes $1/3$ both in (a)
and (b).}
\label{scalingplots}
\end{figure}

\subsection{Scaling analysis of domain growth}

The result in Fig.~\ref{k-t}(a) can be further analyzed in terms of the scaling argument.
In the case of $\gamma=0$, for which the average cancer fraction remains constant
(conserved case), the system exhibits a macrophase separation because such a case 
without any hydrodynamics corresponds to Model B.
Let us denote the steady state characteristic wave number as $k_\infty$. 
In Fig.~\ref{scalingplots}(a), we plot $k_\infty$ as a function of $\gamma$ 
used in Fig.~\ref{k-t}(a).
We find a clear power-law behavior, i.e., $k _\infty \sim \gamma^s$ with $s \approx 0.32$. 
This result suggests that the characteristic wave number obeys the following scaling form
\begin{align}
\langle k (t) \rangle \sim t^{-\alpha} f(\gamma t),
\label{scaling}
\end{align}
where $\alpha$ is the domain growth exponent in the absence of the cancer proliferation effect,
and $f(z)$ is a scaling function with a dimensionless variable $z=\gamma t$.  
A similar scaling hypothesis was successfully used to analyze the phase separation dynamics 
of chemically reactive binary mixtures~\cite{Glotzer94,Christensen96} or that of block copolymer 
melts~\cite{Oono88,Liu89,Bahiana90}.

According to the evaporation-condensation process considered by Lifshitz and Slyozov~\cite{LifshitzBook}, 
the growth exponent should be $\alpha=1/3$ when hydrodynamic interactions are absent. 
This exponent is indeed observed and shown by the dashed line in Fig.~\ref{k-t}(a) before the saturation time.
The asymptotic behavior of the scaling function should be $f(z) \sim {\rm const.}$\ for $z \ll 1$, 
and $f(z) \sim z^{\alpha}$ for $z \gg 1$. 
The latter power-law behavior is required because $\langle k (t) \rangle$ should not depend on time $t$ in the 
steady state. 
Hence, we immediately obtain $k_\infty \sim \gamma^{\alpha}$ and $s=\alpha$.
In Fig.~\ref{scalingplots}(b), we have replotted the quantity $\langle k (t) \rangle t^{1/3}$ as a function 
of $\gamma t$ using all the data in Fig.~\ref{k-t}(a).
The collapse of all the curves demonstrates that our simulation results are in good agreement with the 
above scaling ansatz as long as $\gamma$ is small enough.

In Fig.~\ref{k-t}(b) with full hydrodynamic interactions, the growth exponent in the intermediate stage is 
as large as $\alpha=2/3$ which is much larger than that in Fig.~\ref{k-t}(a). 
However, this result does not obey a simple scaling behavior because of the complicated undershooting 
behaviors.
Here we point out that the value $\alpha=2/3$ was discussed by Furukawa who considered the 
interplay between the inertia of the fluid and the surface energy density~\cite{Furukawa94,OnukiBook}.
This growth exponent was also confirmed by lattice Boltzmann simulations for a critical quench 
of a 2D binary fluid when the viscosity is small and stochastic noise is 
absent~\cite{Osborn95,Gonnella99}. 
Our result cannot be directly compared with theirs because the average composition 
varies with time and also the system exhibits a microphase separation in the late stage. 
However, it is evident from Fig.~\ref{k-t}(b) that a substantial acceleration of phase separation 
takes place in the presence of hydrodynamic flows.

\section{Mechanisms for pattern formation}
\label{sec:hydro}

\subsection{Early stage}

In the early stage of phase separation, when $\phi$ is mostly uniform with small perturbations, we 
are able to analyze the pattern formation by using the amplitude equations method with which the variations of 
$\phi$ and $\mathbf{v}$ are viewed as a group of perturbation waves:
\begin{align}
\phi(\mathbf{r},t) & \approx \langle \phi(t) \rangle +\left[\sum_{\mathbf{q}} \delta\phi_{\mathbf{q}}(t) e^{i\mathbf{q}\cdot\mathbf{r}} +{\rm c.c.}\right], 
\label{phi_AE}\\
\mathbf{v}(\mathbf{r},t) & \approx \sum_{\mathbf{q}} \mathbf{v}_{\mathbf{q}}(t) 
e^{i\mathbf{q}\cdot\mathbf{r}} +{\rm c.c.}, 
\label{velocity_AE}
\end{align} 
where ${\rm c.c.}$\ denotes the complex conjugate and the summation of $\mathbf{q}$ is taken over 
the principal modes of the pattern of interest.

The amplitude equations can be derived by substituting Eqs.~(\ref{phi_AE}) and (\ref{velocity_AE}) into 
Eqs.~(\ref{eq_phic}) and (\ref{eq_velocity}): 
\begin{align}
\frac{d\langle\phi\rangle}{d t} & \approx \gamma\langle\phi\rangle\left(1-\frac{\langle\phi\rangle}{\phi_\infty}\right) 
-\frac{\gamma}{\phi_\infty}\sum_{\mathbf{q}}|\delta\phi_{\mathbf{q}}|^2, \\
\frac{d\delta\phi_{\mathbf{q}}}{d t} & \approx  -i\mathbf{q}\cdot \sum_{\mathbf{q}_1+\mathbf{q}_2 = \mathbf{q}}(\delta\phi_{\mathbf{q}_1}\mathbf{v}_{\mathbf{q}_2}) 
-q^2\mu_{\mathbf{q}} 
\nonumber \\
& +\gamma\delta\phi_{\mathbf{q}}\left(1-\frac{2\langle\phi\rangle}{\phi_\infty}\right)
-\frac{\gamma}{\phi_{\infty}}\sum_{\mathbf{q}_1+\mathbf{q}_2 = \mathbf{q}}(\delta\phi_{\mathbf{q}_1}\delta\phi_{\mathbf{q}_2}), \\
\rho\frac{d \mathbf{v}_{\mathbf{q}}}{d t} & = -\eta q^2\mathbf{v}_{\mathbf{q}} -i\mathbf{q}p_{\mathbf{q}} +i\mathbf{q}\cdot
\boldsymbol{\Sigma}_{\mathbf{q}} -\zeta\mathbf{v}_{\mathbf{q}},
\end{align}
where $q = \vert \mathbf{q} \vert$.
In the above, $\mu_{\mathbf{q}}$, $p_{\mathbf{q}}$ and $\boldsymbol{\Sigma}_{\mathbf{q}}$ are 
the $\mathbf{q}$-th component of the Fourier series of $\mu$, $p$ and $\boldsymbol{\Sigma}$, respectively, 
and are given by
\begin{align}
\mu_{\mathbf{q}} &\approx 
\left[\frac{1}{\langle\phi\rangle}+\frac{1}{1-\langle\phi\rangle} -2\chi +\kappa q^2\right]\delta\phi_{\mathbf{q}}\nonumber\\
&+\frac{1}{2}\left[-\frac{1}{\langle\phi\rangle^2}+\frac{1}{(1-\langle\phi\rangle)^2}\right]
\sum_{\mathbf{q}_1+\mathbf{q}_2=\mathbf{q}} \delta\phi_{\mathbf{q}_1}\delta\phi_{\mathbf{q}_2}\nonumber\\
&+\frac{1}{3}\left[\frac{1}{\langle\phi\rangle^3}+\frac{1}{(1-\langle\phi\rangle)^3}\right]
\sum_{\mathbf{q}_1+\mathbf{q}_2+\mathbf{q}_3 = \mathbf{q}} \delta\phi_{\mathbf{q}_1}\delta\phi_{\mathbf{q}_2}\delta\phi_{\mathbf{q}_3},\\
p_{\mathbf{q}} &= \hat{\mathbf{q}}\cdot\boldsymbol{\Sigma}_{\mathbf{q}}\cdot\hat{\mathbf{q}},\\
\boldsymbol{\Sigma}_{\mathbf{q}} &=  \kappa\sum_{\mathbf{q}_1+\mathbf{q}_2 = \mathbf{q}} 
\left(\mathbf{q}_1\otimes\mathbf{q}_2\right)\delta\phi_{\mathbf{q}_1}\delta\phi_{\mathbf{q}_2}, 
\end{align}
where $\hat{\mathbf{q}} \equiv \mathbf{q}/q$ is the unit vector and $\otimes$ represents the dyadic product.
Then the previous amplitude equations can be simplified as 
\begin{align}
\frac{d\delta\phi_{\mathbf{q}}}{d t} & \approx 
\lambda_1\delta\phi_{\mathbf{q}} 
+\lambda_2\sum_{\mathbf{q}_1+\mathbf{q}_2 = \mathbf{q}}\delta\phi_{\mathbf{q}_1}\delta\phi_{\mathbf{q}_2}\nonumber\\
&+\lambda_3\sum_{\mathbf{q}_1+\mathbf{q}_2+\mathbf{q}_3=\mathbf{q}}\delta\phi_{\mathbf{q}_1}\delta\phi_{\mathbf{q}_2}\delta\phi_{\mathbf{q}_3}\nonumber\\
&-i\mathbf{q}\cdot\sum_{\mathbf{q}_1+\mathbf{q}_2 = \mathbf{q}}(\delta\phi_{\mathbf{q}_1}\mathbf{v}_{\mathbf{q}_2}), 
\label{evolution_phi_k}\\
\rho\frac{d \mathbf{v}_{\mathbf{q}}}{d t} & = -(\eta q^2 +\zeta)\mathbf{v}_{\mathbf{q}}+i\mathbf{q}\cdot\boldsymbol{\Sigma}_{\mathbf{q}}\cdot(\mathbf{I}-\hat{\mathbf{q}}\otimes\hat{\mathbf{q}}),
\label{evolution_vk}
\end{align}
where $\mathbf{I}$ is the unit tensor and the three coefficients in Eq.~(\ref{evolution_phi_k}) are 
given by 
\begin{align}
\lambda_1 &= -q^2\left[\frac{1}{\langle\phi\rangle}+\frac{1}{1-\langle\phi\rangle} -2\chi +\kappa q^2\right]
+\gamma\left(1-\frac{2\langle\phi\rangle}{\phi_\infty}\right),\\
\lambda_2 &= -\frac{q^2}{2}\left[-\frac{1}{\langle\phi\rangle^2}+\frac{1}{(1-\langle\phi\rangle)^2}\right] -\frac{\gamma}{\phi_\infty},\\
\lambda_3 &= -\frac{q^2}{3}\left[\frac{1}{\langle\phi\rangle^3}+\frac{1}{(1-\langle\phi\rangle)^3}\right] <0.
\end{align}

According to the above amplitude equations, it is clear that the hydrodynamic interaction, 
described by the last term in Eq.~(\ref{evolution_phi_k}), is a higher order contribution which does not 
influence the early stage dynamics. 
The linear term $\lambda_1\delta\phi_{\mathbf{q}}$ in Eq.~(\ref{evolution_phi_k}) is independent of  
hydrodynamic interactions and dominates when $\delta\phi_{\mathbf{q}}$ is small.
Therefore, the early stage dynamics must be similar regardless of the values of $\zeta$ as seen in 
Fig.~\ref{patterns2} for $t \leq 10^3$.

According to the time evolution of the velocity in Eq.~(\ref{evolution_vk}), the combination 
$\eta q^2 +\zeta$ controls the decay of the hydrodynamic flow.
This implies that hydrodynamic interactions play a significant role for large length scales. 
Consequently, the flow is suppressed in the early stage when the average wave number 
$\langle k (t) \rangle$ is large, whereas it is strengthened when $\langle k \rangle$ decreases 
as pattern evolves.

\subsection{Late stage}

\begin{figure}[tb]
\centering
\includegraphics[scale=0.6]{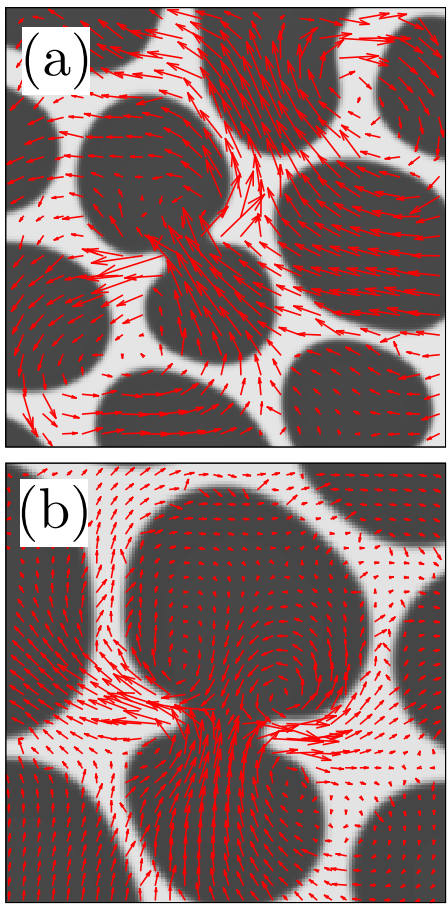}
\caption{
(Color Online) Plots of the velocity field $\mathbf{v}(\mathbf{r}, t)$ shown by the 
arrows at (a) $t=7800$ (system size $200 \times 200$) and (b) $t=9400$
(system size $150 \times 150$) when $\gamma=1\times10^{-3}$ in the presence 
of full hydrodynamic interactions ($\zeta=0$).
The other dimensionless parameters are $\phi_0=0.3$, $\phi_\infty=0.8$, $\chi=2.5$, $\kappa=1$, 
$\rho=0.3$ and $\eta=1.0$.
Both patterns are the closeups of a larger system size simulation as presented by the 
bottom panels of Fig.~\ref{patterns1}.
See also \texttt{SM1.mp4} in the SM.
}
\label{velocity}
\end{figure}

To discuss the late stage dynamics from the viewpoint of hydrodynamic flows, 
we show in Fig.~\ref{velocity}  the velocity field $\mathbf{v}(\mathbf{r}, t)$ together with 
the cancer fraction field $\phi(\mathbf{r}, t)$ in the presence of full hydrodynamic interactions 
($\zeta=0$) at (a) $t=7800$ and (b) $t=9400$
when $\gamma=1\times10^{-3}$ (see also \texttt{SM1.mp4}).
In Fig.~\ref{velocity}(a), a large scale pair of vortices is created; one of them rotates clockwise and 
the other moves counterclockwise. 
Such a flow is triggered by the coalescence of two smaller domains into a larger domain. 
As a result, a strong flow is induced at the neck region of the two merging domains.  
Somewhat later in Fig.~\ref{velocity}(b), on the other hand, a circular flow appears inside a large 
domain. 
Another important feature in this pattern is the existence of a flow along the domain boundaries.
Such a flow sometimes induces a large velocity field in the narrow channel between larger domains.

In the late stage of pattern evolution, the domain structures of the healthy and cancer cells become
relatively robust. 
The values of $\phi$ within healthy-rich and cancer-rich domains are saturated to $\phi \approx 0.145$ 
and $\phi \approx 0.855$, respectively, which correspond to the two free energy minima of 
Eq.~(\ref{freeenergy}) when $\chi=2.5$. 
Once the microstructure is formed, the subsequent evolution of pattern is determined by the competition 
between two different processes; the shape accommodation and the coalescence process.

The shape accommodation results from the movement of interfaces that tends to minimize total 
interfacial energy. 
Therefore, the system energetically favors circular domains and the resultant pattern is the C/H pattern 
composed of circular cancer domains separated by healthy cells. 
On the other hand, the interface is not static due to non-zero net proliferation rate and the coalescence 
occurs when two nearby cancer domains continue to grow and eventually connect each other. 
For larger proliferation rates, the coalescence surpasses the shape accommodation process.
Hence cancer domains get inter-connected and the length scale of pattern increases.
This process leads to a breakdown of the six-fold symmetry of the C/H pattern owing to the 
random connecting processes.

Since the pattern is kinetically controlled by these two processes, the steady state should depend on the 
values of $\gamma$ and $\zeta$, as summarized in Fig.~\ref{diagram}. 
The rate of coalescence process is influenced by the domain growth rate $\gamma$.
The shape accommodation is realized through the mass transportation and it is enhanced by the 
additional hydrodynamics flows across interface, as presented in Fig.~\ref{velocity}.
Thus, when $\gamma$ is as large as $\gamma \approx 5\times 10^{-3}$, the dominating coalescence 
process connects all domains together and transforms the pattern into a uniform cancer cells with few 
healthy spots, corresponding to the H/C patterns.
On the other hand, the shape accommodation process is faster than the coalescence process 
for small $\gamma$ so that the C/H pattern is preserved in the late stage.
The intermediate stripe-like pattern (AB pattern) appears in the steady state when the coalescence 
and shape accommodation processes are comparable.

\section{Summary and discussion}
\label{sec:discussion}

In this paper, we have performed numerical simulations of pattern formation of skin cancers.
In our phase separation model for a binary cellular system, we have taken into account the effects 
of cancer proliferation and hydrodynamic interactions to describe the time evolutions of 
cancer cells. 
As a result of the proliferation effect, the emerging patterns drastically change their structures 
depending on the different stages of the phase separation dynamics.

By controlling the cancer proliferation rate $\gamma$ and the friction coefficient $\zeta$
between dermis and epidermis, we have obtained various types of steady state cancer pattern 
such as a cancer-in-healthy pattern (C/H), a healthy-in-cancer pattern (H/C) and an {locally} 
asymmetric bicontinuous (AB) structure. 
As summarized in Fig.~\ref{diagram}, we have constructed the steady state pattern diagram for different 
combinations of $\gamma$ and $\zeta$ values.
In particular, the C/H patterns obtained for a small proliferation rate and strong hydrodynamic 
interactions (small $\zeta$) and the AB structures obtained for weak hydrodynamic interactions 
(large $\zeta$) might correspond to the globule and the stripe patterns, respectively, in real melanoma 
diagnoses.

For a quantitative analysis, we have calculated the spatially averaged composition of cancer cells,
$\langle \phi (t) \rangle$, and the characteristic length of the cancer patterns, 
$\langle k (t) \rangle$,  as a function of time $t$ (see Figs.~\ref{phi-t} and \ref{k-t}) both in 
the presence and the absence of hydrodynamic interactions. 
We have shown that $\langle \phi (t) \rangle$ and $\langle k (t) \rangle$ depend not only 
on the proliferation rate but also on the strength of hydrodynamic interactions.
Without hydrodynamic flows, we have confirmed in Fig.~\ref{scalingplots} that the scaling 
behavior of the characteristic length is described by the form of Eq.~(\ref{scaling}).
With hydrodynamic flows, on the other hand, the domain growth exponent in the 
intermediate stage was as large as $\alpha=2/3$, showing a pronounced acceleration of the 
microphase separation.

{
First we shall give some numbers for the quantities mentioned in Sec.~\ref{sec:model}C 
to scale length, energy and time that are relevant to skin cancers
(see Eq.~(\ref{dimensionless})).
The typical length scale observed in skin cancer patterns is in the order of $10^{-3}$~m.
According to Fig.~\ref{k-t}, the characteristic wave number in the steady state of our
simulation is $\langle k \rangle a \approx 0.1$ (notice that we recover the 
dimensions of the physical quantities in this Section). 
From these values, we set the unit of length as $a \approx 10^{-5}$~m which corresponds 
to the size of an epidermal cell~\cite{Chatelain2011NJP}. 
Since the interstitial fluid pressure in skin carcinoma was estimated to be roughly  
$\Pi \approx 10^3$~Pa~\cite{Chatelain2011NJP,Jain}, we obtain the typical energy scale as 
$\beta^{-1} \sim \Pi a^3 \approx 10^{-12}$~J that is much larger than the thermal energy.
From the data of the interphase friction~\cite{Chatelain2011NJP,vanKemenade03,Swabb74}, the 3D transport 
coefficient can be evaluated as $L_{\rm 3D} \approx 10^{-15}$~m$^2$$\cdot$Pa$^{-1}$$\cdot$s$^{-1}$. 
With this value, we estimate the typical time scale in our model as 
$a^4\beta/L \sim a^5\beta/L_{\rm 3D} \approx 10^2$~s.
}

{
Having discussed various scales for skin cancers, we can convert the dimensionless parameters
in our simulations to the physical quantities with dimensions. 
For example, the dimensionless time $t / (a^4\beta/L) \approx 10^5$ to reach the steady states 
in Fig.~\ref{k-t} roughly corresponds to $t \approx 10^2$~days which are reasonable for 
cancer spreading.
The choice $\widetilde{\eta} = L \eta/a^2 =1$ in our simulation 
corresponds to $\eta_{\rm 3D} \sim \eta/a \approx 10^5$~Pa$\cdot$s that fits within the 
previously reported viscosity values~\cite{Basan2009,WeiTing2016}.
As for the cancer proliferation rate, the value $\widetilde{\gamma} \sim a^4\beta\gamma/L =10^{-3}$
roughly corresponds to $\gamma \approx 10^{-5}$~s$^{-1} \approx 1$~day$^{-1}$.
This proliferation rate is in agreement with that in the previous reports~\cite{Chatelain2011NJP,Creasey79}. 
Finally, the range of the scaled friction coefficient 
$\widetilde{\zeta}= L \zeta = 10^{-3}$ -- $1$ in our simulation predicts 
$\zeta \approx 10^{7}$ -- $10^{10}$~Pa$\cdot$s$\cdot$m$^{-1}$ and it coincides with 
the range of the friction coefficient in Ref.~\cite{Williamson18}. 
}

Next we discuss the role of cancer proliferation effects on the phase separation dynamics. 
In the conventional Model B describing ordinary macrophase separations, a typical time scale 
is set by the transport coefficient $L$.
In the present model, however, the proliferation rate $\gamma$ in Eq.~(\ref{eq_cancerization2}) 
provides us with additional time scale.
Generally speaking, the phase separation dynamics should be determined by the competition between these 
two time scales.
In our simulation, the initial cancer composition started from $\phi_0=0.3$ and $L$ was much larger than 
$\gamma$.
More precisely, we have chosen the dimensionless number as $a^4\beta\gamma/L \approx 10^{-3}$
in the simulations (see Eq.~(\ref{dimensionless})). 
Hence the compositional instability for the phase separation, that is governed by $L$, takes place before 
the average composition $\langle \phi(t)\rangle$ increases with the rate $\gamma$.

As shown in Fig.~\ref{patterns1}, the cancer domains appear as a result of unstable concentration 
fluctuations, and they form C/H patterns for $\langle \phi (t) \rangle<0.5$ in the early stage.
In the late stage, the initial C/H pattern continues to remain for smaller $\gamma$ values, while it 
transforms into the H/C pattern for larger $\gamma$ values.  
When the quantity $a^4\beta\gamma/L $ is much larger and becomes close to unity, the system 
always exhibits the H/C pattern because the average composition will be immediately saturated at 
a larger value $\langle \phi (t) \rangle>0.5$ before the system undergoes a phase separation.
Hence the cancer proliferation significantly affects the microstructures of cancer patterns.

In the present work, we have considered a 2D system composed of cancer and healthy cells whose 
compositions evolve in time due to the cancer proliferation effect. 
Although a similar model was proposed by Chatelain \textit{et al.}~\cite{Chatelain2011NJP,Chatelain2011JTB}, 
the main difference in our work is that the effects of hydrodynamic interactions are explicitly taken 
into account. 
Moreover, the strength of hydrodynamic interactions can be controlled by changing the friction
coefficient $\zeta$.
When hydrodynamic interactions are fully present, the C/H patterns continue to remain even in the 
late stage when $\langle \phi (t) \rangle >0.5$ (see bottom panels in Fig.~\ref{patterns1}
and \texttt{SM1.mp4}).
Such a transient pattern was not observed in the previous study by Chatelain 
\textit{et al.}~\cite{Chatelain2011NJP,Chatelain2011JTB}.

Alternatively, Chatelain \textit{et al.}\ took into account the diffusion of nutrient (oxygen) concentration 
chosen as an additional variable~\cite{Chatelain2011NJP,Chatelain2011JTB}.
Accordingly, they employed a diffusion equation for the nutrient concentration with a 
source term.
In their model, the cell-nutrient interaction defines a typical diffusive length that controls the 
saturation of growing domains.
In our model, we did not consider such a coupling to the diffusion of nutrients from an outer 
environment, but simply used the logistic growth model to describe the cancer proliferation
(see Eq.~(\ref{eq_cancerization2})). 
As mentioned before, this simplification is  justified when the cancer composition is proportional to 
the nutrient concentration.

We have assumed that dermal/epidermal boundary is flat and the epidermal layer was modeled 
as a 2D fluid. 
However, the structure of dermis and epidermal can affect the cell differentiation and also the 
cancer pattern formation.
For example, Balois \textit{et al.}\ considered melanin transport in epidermis and showed that 
it is influenced by the dermal/epidermal shape~\cite{Balois2014JRSI}.
Such a geometrical effect of basal layer will be considered in our future study by taking into 
account the hydrodynamic interaction.

{
Cates \textit{et al.}\ argued that the appearance of an arrested 
phase separation in bacterial colonies can be explained only by considering a local density-dependent 
motility and the birth/death of bacteria~\cite{Cates2010}.
In their work, the competition between the effects of birth/death and diffusion leads to a typical 
length scale beyond which domain coarsening does not occur.
The obtained patterns of 2D simulation indeed show droplets of the high-density phase dispersed in a 
continuous low-density phase at large times~\cite{Cates2010}.
Such a situation is very reminiscent to the results of our model in the absence hydrodynamic 
interactions (either C/H or H/C pattern).
On the other hand, we have shown that hydrodynamic interactions affect not only the 
steady state patterns but also the transient patterns.
}

In Sec.~\ref{sec:model}A, we have mentioned that the logistic growth of cancer cells in 
Eq.~(\ref{eq_cancerization2}) can stem from the mechanical coupling effect that is 
controlled by the homeostatic pressure~\cite{Basan2009}.
Ranft \textit{et al.}\ discussed the propagation of an interface between two different cell 
populations when the homeostatic pressures of two cell types are different~\cite{Ranft14}.
Taking into account both substrate friction and hydrodynamic interactions,  Podewitz 
\textit{et al.}\ performed mesoscopic simulations to investigate interface dynamics of 
competing tissues~\cite{Podewitz16}.
They showed that the propagation velocity of the interface is proportional to the homeostatic 
stress difference.
Recently, Williamson and Salbreux studied the stability and roughness of such a propagating 
interface~\cite{Williamson18}.
In these studies, however, the formation of microstructures of cancer cells, such as 
dots or stripes, has not been investigated.

As mentioned before, our model can reproduce clinically observed globule and stripe patterns in melanoma.  
The C/H patterns tend to appear when the proliferation rate is small and the hydrodynamic interactions 
are strong.
By contrast, the stripe patterns,  which are often found in human palms or soles, tend to appear 
when hydrodynamic interactions are absent.
In reality, palms and soles contain a thick stratum corneum and an unique cell layer called ``stratum 
lucidum" which has a finite stiffness. 
Such a stiffness may reduce hydrodynamic interactions and results in the formation of stripe patterns.

Our model suggests that the proliferation and invasion of cancer cells in superficial spreading melanoma can be 
predicted by observing the epidermis using dermoscopy.
Melanoma cells migrate horizontally in the epidermis in the initial stage of tumor development, during which the 
clinical staging is described by ``Clark's level" and ``Breslow's depth"~\cite{JamesBook}.
In its staging, the diffusion range and the cell spreading pattern of melanoma cells are the most important 
measures for making prognostic predictions, such as the five-year patient survival rate~\cite{BologniaBook,Tas2017}.
The present work presents objective diagnostic indicators and methodologies for making prognostic predictions 
for these patients that can be verified by dermoscopic image data.
We expect that our work will be applied to the development and evaluation of future clinical diagnosis.

\section*{Acknowledgements}
We thank R.\ Okamoto, K.\ Yasuda, T.\ Kato, and R.\ Kurita for useful discussions.
T.H.\ thanks the hospitality of National Tsing Hua University and National Central University 
where part of this research was conducted under the Co-Tutorial Program.
T.H.\ acknowledges the support by Grant-in-Aid for JSPS Fellows (Grant No.\ 17J01643)
from the Japan Society for the Promotion of Science (JSPS).
K.-A.W.\ and M.-W.L.\ acknowledge the support of the Ministry of Science and Technology, 
Taiwan (Grant No.\ MOST 105-2112-M-007-031-MY3).
K.-A.W., M.-W.L., and H.-Y. C.\ thank the support from National Center for Theoretical Sciences, Taiwan.
S.K.\ acknowledges the support by Grant-in-Aid for Scientific Research (C) (Grant No.\ 18K03567) 
from the JSPS.



\begin{thebibliography}{99}

\bibitem{MorphogenesisBook}
Edited by V. Capasso, M. Gromov, and A. Harel-Bellan, 
\textit{Pattern Formation in Morphogenesis}
(Springer, Heidelberg, 2013).

\bibitem{Komarova2005}
N. L. Komarova,
Curr. Opin. Oncol. \textbf{17}, 39 (2005).

\bibitem{Pazdziorek2014}
P. R. Pa\'zdziorek,
Bull. Math. Biol. \textbf{76}, 1642 (2014).

\bibitem{Basan2009}
M. Basan, T. Risler, J.-F. Joanny, X. Sastre-Garau, and J. Prost, 
HFSP J. \textbf{3},  265 (2009).

\bibitem{Ranft2010}
J. Ranft, M. Basan, J. Elgeti, J.-F. Joanny, J. Prost, and F. J\"{u}icher,
Proc. Natl. Acad. Sci. USA \textbf{107}, 20863 (2010).

\bibitem{Basan11}
M. Basan, J. Prost, J.-F. Joanny, and J. Elgeti,
Phys. Biol.  \textbf{8},  026014 (2011).

\bibitem{Brochard2012}
D. Gonzalez-Rodriguez, K. Guevorkian, S. Douezan, and F. Brochard-Wyart,
Science \textbf{338}, 910 (2012).

\bibitem{Kumar2009}
S. Kumar and V. M. Weaver,
Cancer Metastasis Rev. \textbf{28}, 113 (2009).

\bibitem{Deisboeck2011}
T. S. Deisboeck, Z. Wang, P. Macklin, and V. Cristini,
Annu. Rev. Biomed. Eng. \textbf{13}, 127 (2011).

\bibitem{Rodriguez2013}
I. A. Rodriguez-Brenes, N. L. Komarova, and D. Wodarz,
Trends Ecol. Evol. \textbf{28}, 597 (2013).

\bibitem{Liedekerke2015}
P. Van Liedekerke, M. M. Palm, N. Jagiella, and D. Drasdo, 
Comp. Part. Mech. \textbf{2}, 401 (2015).

\bibitem{Chatelain2015}
C. Chatelain and M. Ben Amar,
Eur. Phys. J. Plus \textbf{130}, 176 (2015). 

\bibitem{Chatelain2011NJP}
C. Chatelain, T. Balois, P. Ciarletta, and M. Ben Amar,
New J. Phys. \textbf{13}, 115013 (2011). 

\bibitem{Chatelain2011JTB}
C. Chatelain, P. Ciarletta, and M. Ben Amar,
J. Theo. Bio. \textbf{290}, 46 (2011). 

\bibitem{Balois2014}
T. Balois and M. Ben Amar,
Sci. Rep. \textbf{4}, 3622 (2014).

\bibitem{Balois2014JRSI}
T. Balois, C. Chatelain, and M. Ben Amar,
J. R. Soc. Interface \textbf{11}, 20140339 (2014). 

\bibitem{HamleyBook}
I. W. Hamley,
\textit{The Physics of Block Copolymers} 
(Oxford University Press, Oxford, 1998).

\bibitem{Hohenberg77}
P. C. Hohenberg and B. I. Halperin, 
Rev. Mod. Phys. \textbf{49}, 435 (1977).

\bibitem{LubenskyBook}
P. M. Chaikin and T. C. Lubensky, 
\textit{Principles of Condensed Matter Physics} 
(Cambridge University Press, Cambridge, 1995).

\bibitem{OnukiBook}
A. Onuki, 
\textit{Phase Transition Dynamics} 
(Cambridge University Press, Cambridge, 2002).

\bibitem{Cates2010}
M. E. Cates, D. Marenduzzo,  I. Pagonabarraga, and J. Tailleur,
Proc. Natl. Acad. Sci. USA \textbf{107}, 11715 (2010).

\bibitem{Steinberg63}
M. S. Steinberg, 
Science \textbf{141}, 401 (1963).

\bibitem{Steinberg96}
M. S. Steinberg, 
Dev. Biol. \textbf{180}, 377 (1996).

\bibitem{He2014}
B. He, K. Doubrovinski, O. Polyakov, and E. Wieschaus,
Nature \textbf{508}, 392 (2014).

\bibitem{WeiTing2016}
W.-T. Yeh and H.-Y. Chen,
Phys. Rev. E \textbf{93}, 052421 (2016).

\bibitem{WeiTing2018}
W.-T. Yeh and H.-Y. Chen,
New J. Phys. \textbf{20}, 053051 (2018).

\bibitem{Siggia79}
E. D. Siggia, 
Phys. Rev. A \textbf{20}, 595 (1979).

\bibitem{Kendon01}
V. M. Kendon, M. E. Cates, I. Pagonabarraga, J.-C. Desplat, and P. Bladon,  
J. Fluid Mech. \textbf{440}, 147 (2001).

\bibitem{Binder74}
K. Binder and D. Stauffer, 
Phys. Rev. Lett. \textbf{33}, 1006 (1974).

\bibitem{LifshitzBook}
E. M. Lifshitz and L. P. Pitaevskii, 
\textit{Physical Kinetics}
(Pergamon Press, Oxford, 1981).

\bibitem{Glotzer94}
S. C. Glotzer, D. Stauffer, and N. Jan,
Phys. Rev. Lett. \textbf{72},  4109 (1994).

\bibitem{Christensen96}
J. J. Christensen, K. Elder, and H. C. Fogedby,
Phys. Rev. E \textbf{54}, R2212 (1996).

\bibitem{Oono88}
Y. Oono and M. Bahiana,
Phys. Rev. Lett. \textbf{61}, 1109 (1988).

\bibitem{Liu89}
F. Liu and N. Goldenfeld, 
Phys. Rev.  A \textbf{39}, 4805 (1989).

\bibitem{Bahiana90}
M. Bahiana and Y. Oono, 
Phys. Rev.  A \textbf{41}, 6763 (1990).

\bibitem{Jain}
R. K. Jain,
Cancer Res. \textbf{47}, 3039 (1987).

\bibitem{Montel11}
F. Montel, M. Delarue, J. Elgeti, L. Malaquin, M. Basan, T. Risler, B. Cabane, D. Vignjevic, 
J. Prost, G. Cappello, and J.-F. Joanny,
Phys. Rev. Lett. \textbf{107},  188102 (2011).

\bibitem{Montel12}
F. Montel, M. Delarue, J. Elgeti, D. Vignjevic, G. Cappello, and J. Prost,
New J. Phys. \textbf{14}, 055008  (2012).

\bibitem{Alessandri13}
K. Alessandri, B. R. Sarangi, V. V. Gurchenkov, B. Sinha, T. R. Kie{\ss}ling, 
L. Fetler, F. Rico, S. Scheuring, C. Lamaze,  A. Simon, S. Geraldo, D. Vignjevi\'{c}, 
H. Dom\'{e}jean, L. Rolland,  A. Funfak, J. Bibette, N. Bremond, and P. Nassoy, 
Proc. Natl. Acad. Sci. USA \textbf{110},  14843 (2013).

\bibitem{Delarue13}
M. Delarue, F. Montel, O. Caen, J. Elgeti, J.-M. Siaugue, D. Vignjevic, J. Prost, J.-F. Joanny, and G. Cappello,
Phys. Rev. Lett. \textbf{110},  138103 (2013).

\bibitem{Tiribocchi15}
A. Tiribocchi, R. Wittkowski, D. Marenduzzo and M. E. Cates,
Phys. Rev. Lett. \textbf{115}, 188302 (2015).

\bibitem{Puri97}
S. Puri, A. J. Bray, J. L. Lebowitz, 
Phys. Rev. E \textbf{56}, 758 (1997).

\bibitem{Ahluwalia99}
R. Ahluwalia,
Phys. Rev. E \textbf{59}, 263 (1999).

\bibitem{Wise10}
S. M. Wise, J. S. Lowengrub, H. B. Frieboes, and V. Cristini, 
J. Theo. Bio. \textbf{253}, 524 (2008). 

\bibitem{DoiBook}
M. Doi, 
\textit{Soft Matter Physics} 
(Oxford University, Oxford, 2013).

\bibitem{Harlow65}
F. H. Harlow and J. E. Welch,
Phys. Fluids \textbf{8},  2182 (1965).

\bibitem{Shinozaki93}
A. Shinozaki and Y. Oono,
Phys. Rev. E \textbf{48}, 2622 (1993).

\bibitem{Furukawa94}
H. Furukawa, 
Physica \textbf{204A}, 237 (1994).

\bibitem{Osborn95}
W. R. Osborn, E. Orlandini, M. R. Swift, J. M. Yeomans, and J. R. Banavar,
Phys. Rev. Lett. \textbf{75}, 4031 (1995).

\bibitem{Gonnella99}
G. Gonnella, E. Orlandini, and J. M. Yeomans,
Phys. Rev. E \textbf{59}, R4741 (1999).

{
\bibitem{vanKemenade03}
P. M. van Kemenade, J. M. Huyghe, and L. F. A. Douven, 
Porous Media \textbf{50}, 93 (2003).
}

{
\bibitem{Swabb74}
E. A. Swabb, J. Wei, and P. M. Gullino, 
Cancer Res. \textbf{34}, 2814 (1974).
}

{
\bibitem{Creasey79}
A. A. Creasey, H. S. Smith, A. J. Hackett, K. Fukuyama, W. L. Epstein, and S. H. Madin,
In Vitro \textbf{15}, 342 (1979).
}

\bibitem{Williamson18}
J. J. Williamson and G. Salbreux,
Phys. Rev. Lett. \textbf{121}, 238102  (2018).

\bibitem{Ranft14}
J. Ranft, M. Aliee, J. Prost, F. J\"{u}licher, and J.-F. Joanny,
New J. Phys.  \textbf{16},  035002 (2014).

\bibitem{Podewitz16}
N. Podewitz, F. J\"{u}licher, G. Gompper, J. Elgeti,
New J. Phys. \textbf{18}, 083020  (2016).

\bibitem{JamesBook}
W. D. James, T. G. Berger, and D. M. Elston,
\textit{Andrews'  Diseases  of  the  Skin: Clinical  Dermatology}
(Saunders Elsevier, Philadelphia, 2006). 

\bibitem{BologniaBook}
J. L. Bolognia, J. L. Jorizzo, and R. P. Papini,
\textit{Dermatology}
(Mosby, St. Louis, 2007).

\bibitem{Tas2017}
F. Tas and K. Erturk,
Mol. Clin. Oncol. \textbf{7}, 1083 (2017).

\end{thebibliography}
\end{document}